\documentclass{elsart} 

\usepackage{graphicx}

\usepackage{amssymb}

\begin{document}

\begin{frontmatter}

\title{Density and conformation with relaxed substrate, bulk, 
and interface in electrophoretic deposition of polymer chains}

\author{Frank W. Bentrem, Jun Xie, and R. B. Pandey}

\address{Department of Physics and Astronomy, The University of Southern 
Mississippi,
Hattiesburg, MS 39406-5046}

\begin{abstract}

Characteristics of relaxed density profile and conformation of 
polymer chains are studied by a Monte Carlo simulation on a 
discrete lattice in three dimensions using different segmental 
(kink-jump $K$, crank-shaft $C$, reptation $R$) dynamics.
Three distinct density regimes, substrate,
bulk, and interface, are identified. With the $KC$ segmental 
dynamics we find that the substrate coverage grows with a power-law,
$d_s \propto t^{\gamma}$ with a field dependent nonuniversal
exponent $\gamma = 0.23 + 0.7 E$. The bulk volume fraction $d_b$
and the substrate polymer density ($d_s$) increases exponentially 
with the field ($d_b \propto E^{0.4}$, $d_s \propto E^{0.2}$)
in the low field regime. The interface polymer density $d_f$
increases with the molecular weight. With the $KCR$ segmental 
dynamics, bulk and substrate density decreases linearly with the
temperature at high temperatures. The bulk volume fraction is found
to decay with the molecular weight, $d_b \propto L_c^{-0.11}$.
The radius of gyration remains Gaussian in all density regions.

\end{abstract}


\end{frontmatter}

\section{Introduction}

One of the common processes to design polymeric materials (including
composites) and to coat surfaces \cite{brewer73,wool} is to deposit 
polymer constituents on a desirable substrate and equilibrate. A driving 
(biased) field 
caused by electric field, pressure gradient, gravitation, spinning 
processes, etc., is generally used to direct the deposition and 
control the morphological evolution. In DNA gel electrophoresis 
for example, the chain macromolecules are driven by an electric 
field \cite{zimm,chu}. The flow of DNA 
molecules through pores and their accumulation at the pore boundaries 
along the flow direction depend on the magnitude of the flow field, 
molecular weight, and porosity (i.e., the size of the pores), etc.
As the polymer chains deposit on an impenetrable substrate, the 
polymer density grows and an interface develops 
\cite{fp1,fp2,fp3,bpf1}. The interface width 
defined by the longitudinal fluctuation in density profile is a 
measure of the roughness of the growing surface
\cite{barabasi,family}. 
Enormous efforts have been devoted to understanding the conformation
and density profiles at/near the surface; the list of references
is too large to cite them all, but we provide here a few examples
\cite{wang91,sen95,forsman96,dobrynin01,collins94,pmb1} without 
any preference. However, studies of the interface growth, 
conformation and density profile for the
electrophoretic deposition of polymer chains on the impenetrable
substrate \cite{fp1,fp2,fp3,bpf1} are very limited. 
The characteristics of the density profile, growth of the interface 
width, and roughness depend on the deposition process, parameters
such as driving field, temperature, and molecular weight, and the
relaxation procedures.

Since the polymer chains, driven by a field, deposit on an 
impenetrable substrate/wall, their conformational phase space 
at the substrate is constrained. 
The polymer density grows from the substrate as the deposition 
proceeds which leads to evolution of the bulk region adjacent to
the substrate. 
Distinct regimes emerge from initiating growth at the substrate 
toward the source at the opposite end from where the polymer chains
are released: substrate, bulk, interface, and dilute solution.
Movement of chain segments, their relaxation, and
degree of entanglement may vary from one regime to another.
Parameters such as temperature, molecular weight, field, and 
procedure such as segmental dynamics and relaxation may also
affect the characteristics of these regimes. Therefore, it would
be interesting to study how the polymer density varies from
one region to the next and how it depends on these parameters. 

While it is difficult to monitor and control deposition process,
density profile, polymer conformation, etc. at the molecular
scales in the laboratory, computer simulation experiments 
can be implemented to examine these properties with simplified 
models \cite{binder95}.
Making quantitative predictions by analytical theories is severely
limited in such a complex system involving different phases of
material, interfaces, and relaxation at various scales. Therefore,
computer simulation experiments may be useful to gain insight into
some of the difficult issues such as prediction of density profile
and its dependence on molecular weight, field and temperature.
In recent articles, we have reported our analysis of the interface
growth and its scaling \cite{fp2,fp3,bpf1,bentrem01}. 
Using the same computer simulation model on a discrete lattice 
we would like to study characteristics of the density profile
and conformation of chains in this paper.
Model is presented in the next section followed by results and
discussion in section 3 with a conclusion at the end.

\section{Model and Method}
We consider a discrete lattice of size $L_x \times L \times L$ with 
a large aspect ratio $L_x/L$. Coarse-grained polymer chains,
each of length $L_c$ (i.e., ($L_c+1$) nodes connected by bonds)
are generated on trails of random walks along the lattice with 
excluded volume constraints toward the source end ($x = 1$). 
An external field $E$ drives the chains toward the substrate
(impenetrable wall) at $x = L_x$.
The change in energy due to the field 
$$\Delta U = E \cdot \Delta x, \eqno (1)$$
where $\Delta x$ is the displacement of the chain node along $x$ 
direction.
We also consider a nearest-neighbor polymer-polymer repulsive and 
polymer-wall attractive interaction,
$$U = J \sum_{ij} \rho_i \rho_j, \eqno (2)$$
where $J = 1, \rho_i = 1$ if the site $i$ is occupied by the polymer
node, $0$ if the site $i$ is empty, and $-1$ if the empty site $i$
is on the substrate. The summation is restricted to nearest neighbor
sites. Chains are released at a constant rate (typically one every
hundred time steps) and moved with the Metropolis algorithm
\cite{metropolis53} using a 
combination of segmental movements \cite{binder95}, i.e., kink-jump, 
crankshaft \cite{verdier62}, and slithering snake (reptation) 
dynamics.  
Attempt to move each chain node once is defined as one Monte Carlo 
step (MCS). The simulation is performed for a relatively large 
number of time steps for a sufficiently large number of independent 
samples to obtain a reliable estimate of the averaged physical 
quantities.

Polymer chains move from the source end toward the impenetrable
substrate at the opposite end by the field ($E$) using different
segmental dynamics. Kink jump dynamics involves movement of 
only one node and it is relatively slow and localized.
Therefore, it takes much longer to relax the polymer chains. Crank-shaft 
involves movements of two nodes simultaneously, and it is
faster than kink-jump but still localized to small part of chain's 
segment. Slithering-snake (\lq\lq reptation") dynamics, on the other hand,
involves global motion of the whole chain by altering conformations
at the end segments. Perhaps, a combination of these and other
modes of segmental movements are desirable in many polymer 
simulations. We will, however, concentrate here with combinations of
three segmental dynamics, ($i$) kink-jump ($K$), ($ii$) kink-jump
and crank-shaft ($KC$), and ($iii$) kink-jump, crank-shaft, and
reptation ($KCR$). Apart from the segmental dynamics,
a relaxation procedure \cite{bentrem01}is also implemented to 
equilibrate the polymer conformation and density. If we continue 
to deposit polymer chains and monitor the evolution of the density 
and conformational profiles, interface width, etc., we reach a 
steady-state where these quantities approach their steady-state 
(constant) value in asymptotic time steps \cite{fp2,fp3,bpf1}. 
However, if we stop releasing new chains after
an appropriate, i.e., desirable growth while allowing chains to
execute their segmental moves, we find \cite{bentrem01} a 
considerable change in the interface width (i.e., the roughness). 
We have recently reported \cite{bentrem01}
scaling behaviors of the \lq\lq relaxed" interface width and pointed out
the differences from their \lq\lq steady-state" values.
In this paper, we present our results  of relaxed data for the
density profile and conformation.

\section{Result and Discussion}

Most of our simulations are performed on a $100 \times 40 \times 40$
lattice with the size of polymer chains $L_c = 10-100$ at various
temperatures and fields. A number of independent samples $N_s = 10-20$ are
used to average the physical quantities. Different lattice sizes are also
used to check for the severe finite size effect and no significant
difference is observed in the qualitative behavior of the physical
quantities reported here. We have already reported the growth and scaling
of the interface width including roughness in a series of papers in recent
years \cite{fp2,fp3,bpf1,bentrem01} by analyzing steady-state and
relaxed interface width. In this article we concentrate on the
characteristics of the density in various regimes. Our data on the
conformation of chains in these regimes have relatively large fluctuations
and therefore, their qualitative analysis will be limited.

Some studies for polymer melts near an impenetrable wall
\cite{fp1,wang91,sen95,pmb1} have indicated an oscillation in the
density profile near the wall. Other studies have shown a parabolic
\cite{milner88a,milner88b} or hyperbolic \cite{dobrynin99} decay in
density with distance from the wall.
For our simulations relaxed density profiles with $K$, $KC$, and $KCR$ 
dynamics are presented
in Figs. 1-3, respectively, at the temperature $T=1$ for different
driving fields. Even though the chain lengths are different 
($L_c = 39, 39, 50$), the qualitative forms of their density
profiles should be comparable for the following reason.
We have examined the relaxed density profiles with different
chain lengths $L_c = 10 - 100$ and have found no 
significant difference
in their density profile (independent of chain length) for any 
combination of the segmental
dynamics we have studied here. A careful examination of these
profiles (Figs. 1-3), however, reveals some differences due to
segmental dynamics. Figure 4 shows an example for such a comparison
at $T = 1, E=0.2$ with $K$, $KC$, and $KCR$ segmental dynamics.
We see that the polymer density in most regions (particularly at
the substrate and in the bulk) is highest with the $KCR$ dynamics and
lowest with the $K$ dynamics as it is slowest in relaxing the chains
and is restricted to very short range local moves.
Reptation (slithering-snake) movements in $KCR$ dynamics involves 
long range fast movements which helps equilibrating the polymer
density better than the short-range moves with $K$ and $KC$
segmental dynamics. 

Obviously, there is a profound effect of the field
in orchestrating the shapes of the density profiles. A more specific
examination will follow later. It is worth pointing out that we
have not observed any sign of oscillations in density profiles as
observed with the steady-state density profiles with a combination
of kink-jump and reptation dynamics \cite{fp2}. Further, there is clogging 
with
the kink-jump segmental dynamics alone particularly at higher
values of field ($E \le 0.6$, see Fig.~1). With clogging, the rate
of polymer deposition on the substrate is hard to monitor and
different from that without clogging. Such a problem with the
$K$ segmental dynamics is also reported in study of the interface
growth and its scaling. In the following we restrict most of our
analysis to $KC$ and $KCR$ dynamics where the clogging is not
a problem for the range of parameters we have explored here.
Thus, mixing different segmental
movements along with relaxation (without additional deposition) is
important in relaxing the polymer chains, avoiding the clogging,
equilibrating the density distribution, and reducing the probability 
of layering in density profile of driven polymer deposition.

Figure 5 shows the density profiles at different temperatures with
$KC$ and $KCR$ segmental dynamics at a fixed field $E=0.5$. Again we 
see the differences due to segmental dynamics along with the effect
of temperature, which is revisited later.

Let us examine the characteristics of density in each region, i.e.,
substrate ($x=L_x$), bulk (between substrate and the interface),
interface, and dilute solution (fluid), of the density profile.
Polymer density in bulk is referred to as the volume fraction and
density at the substrate as the coverage, the fraction of sites
covered by the polymer. Figure 6 shows the growth of coverage with
$KC$ segmental dynamics. Note that the coverage, $d_s$, grows rather
fast initially followed by a slow growth before reaching its 
saturation, the asymptotic value also known as jamming coverage.
It is worth pointing out that the approach to
jamming coverage for objects of different shapes and sizes has been
notoriously slow and extensively studied \cite{feder,pomeau,swendsen,ziff,dickman,viot,becklehimer,wang}.
While the substrate coverage in steady-state polymer deposition 
shares the slow approach to the asymptotic (jamming) limit, equilibration
by relaxing chains enhances the probability of approaching the 
jamming limit substantially. In the pre-asymptotic regime, growth
of the coverage seems to grow with a power law (see Fig.~7),
$$d_s \propto t^{\gamma}, \eqno(3) $$
with an exponent $\gamma$. The coverage growth exponent $\gamma$
is found to depend on the field (see Fig.~8),
$$\gamma = 0.23 + 0.7 E, \eqno(4)$$ 
and is therefore, non-universal.

Relaxed densities in the bulk, i.e., the volume fraction $d_b$ and
the substrate coverage $d_s$ depend on the field in low
field regime ($E \le 0.5$) as seen in Fig.~9. The volume fraction
and the coverage increase with the field exponentially and become 
constant at their maximum values at high fields.
The polymer density at the interface $d_f$, on the other hand,
is found to increase with the field with a power law (see Fig 10).
Best fits of our data suggest,
$$d_f \propto E^{0.3}, \, d_b \propto E^{0.4}, 
\, d_s \propto E^{0.2}. \eqno(5)$$
The response of polymer density to field in the low field regime
seems to be stronger with the relaxed system here than the 
corresponding response with a steady-state simulation.

The interface polymer
density also seems to increase with the molecular weight.
Best fit of our data in Fig.~11 leads to,
$$d_f = 0.4 + 0.0002 L_c \eqno(6)$$
It is worth mentioning that a similar trend for the dependence
of volume fraction and coverage on the field is also observed with
the kink-jump dynamics except the range of field for
the growth of density is reduced. 

We have also analyzed the temperature dependence of polymer densities
in these regions (substrate, bulk, interface). 
Generally, the density decreases with the temperature. The substrate 
coverage and the bulk volume fraction show a larger rate of decay 
than the polymer density at the interface.  
Volume fraction in bulk shows the largest variation with the 
temperature for the $KCR$ segmental dynamics as shown in Fig.~12.
Best fits of our data in high temperature regime show,
$$d_b = 1.02 - 0.015 T, \,d_s = 1.00 - 0.0013 T \eqno(7)$$ 
Since the conformation of polymer chains at the substrate
and in the bulk are more restricted due to the presence of neighboring
chains and their entanglement, increasing the temperature further reduces
the conformational entropy. Thus we are able to characterize the 
dependence of volume fraction and substrate coverage on field,
temperature and the molecular weight.

The bulk density is found to decay with the molecular weight as shown
in Fig.~13,
$$d_b \propto L_c^{-0.011} \eqno(8)$$
Note that the decay of bulk density with the molecular weight shows 
a similar pattern with steady state simulation with $KCR$ segmental
dynamics. 
Exponential decay is observed with a steady-state simulation 
involving reptation and kink-jump. 
We know that it is difficult to pack many chains in a limited
volume due to steric hindrance. The larger the chain length the fewer the
chains are needed to reach the jamming limit. The percolation threshold
\cite{stauffer} of chains and its jamming volume fraction have been 
found \cite{becklehimer} to decay 
with the chain length in the percolation of chains. 
Our result seems consistent with the percolation of chains.

We have also analyzed the conformation of polymer chains.
In general, our data for the conformational profiles are
more fluctuating than corresponding density profiles.
Therefore, at present, we are unable to draw more meaningful 
conclusion that what is presented here. 
Figure 14 shows the variation of the radius of gyration of chains
at the interface, substrate, and in the bulk, with the 
chain length on a log-log scale. We see that the transverse 
component of $R_{g}$ scales with the molecular weight with, 
$$R_{gt} \propto L_c^{\nu_t}, \, \nu_t \sim 0.6 \eqno(9).$$ 
The total radius of gyration shows a Gaussian scaling, 
$$R_g \propto L_c^{1/2}. \eqno(10)$$ 
The longitudinal component saturates at higher
molecular weight as chains compress.

\section{Conclusion}

Electrophoretic deposition of polymer chains on an impenetrable
substrate is examined by a computer simulation model on a discrete
lattice. A combination of different segmental dynamics is used
to relax the polymer chains and monitor the density profile. 
We find that the relaxation of polymer chains, density, and the 
interface width depend on the segmental dynamics, molecular weight, 
field, and temperature. The combination of slow-to-fast segmental 
dynamics is highly desirable to equilibrate our system effectively 
in our observation time.

Different spatial regions of the material evolved from the polymer
deposition, substrate, bulk, and the interface are identified and
their physical properties are analyzed. 
We are able to make a number of quantitative predictions regarding
the dependence of relaxed density and conformation of chains in 
these regions based on our data. We have found that there is very
little (negligible) effect of molecular weight on the the relaxed
density profiles both with $KC$ and $KCR$ segmental dynamics.
With the $KC$ segmental dynamics, we find that the substrate coverage
in the pre-saturation regime grows with a power law with a 
nonuniversal exponent $\gamma$ (Eqs. 3 and 4). Bulk volume fraction
($d_b$) and the jamming substrate coverage grow exponentially with
the field (Eq. 5). The interface polymer density ($d_f$) increases
linearly with the molecular weight (Eq. 6).

With the $KCR$ dynamics, we are able to attain better defined
distinct density regions with relatively less fluctuations.
Both bulk volume fraction and substrate coverage are found to
decay linearly with the temperature in the high-temperature
regime. The bulk density decays with the molecular weight with
a power law exponent 0.11 (Eq. 8).

The radius of gyration exhibits a Gaussian conformation in almost
all regions. While the transverse component of the radius of
gyration scales with an exponent (0.6) similar to that of a
self-avoiding walk, the longitudinal components seem to saturates
with higher molecular weight. Further large scale simulations are
needed to clarify the conformational profile. 
\bigskip

\parindent 0pt
{\bf Acknowledgments}: 
We acknowledge partial support from a DOE-EPSCoR grant.

\newpage
\begin{figure}[hbt]
\begin{center}
\includegraphics[angle=-00,scale=0.6]{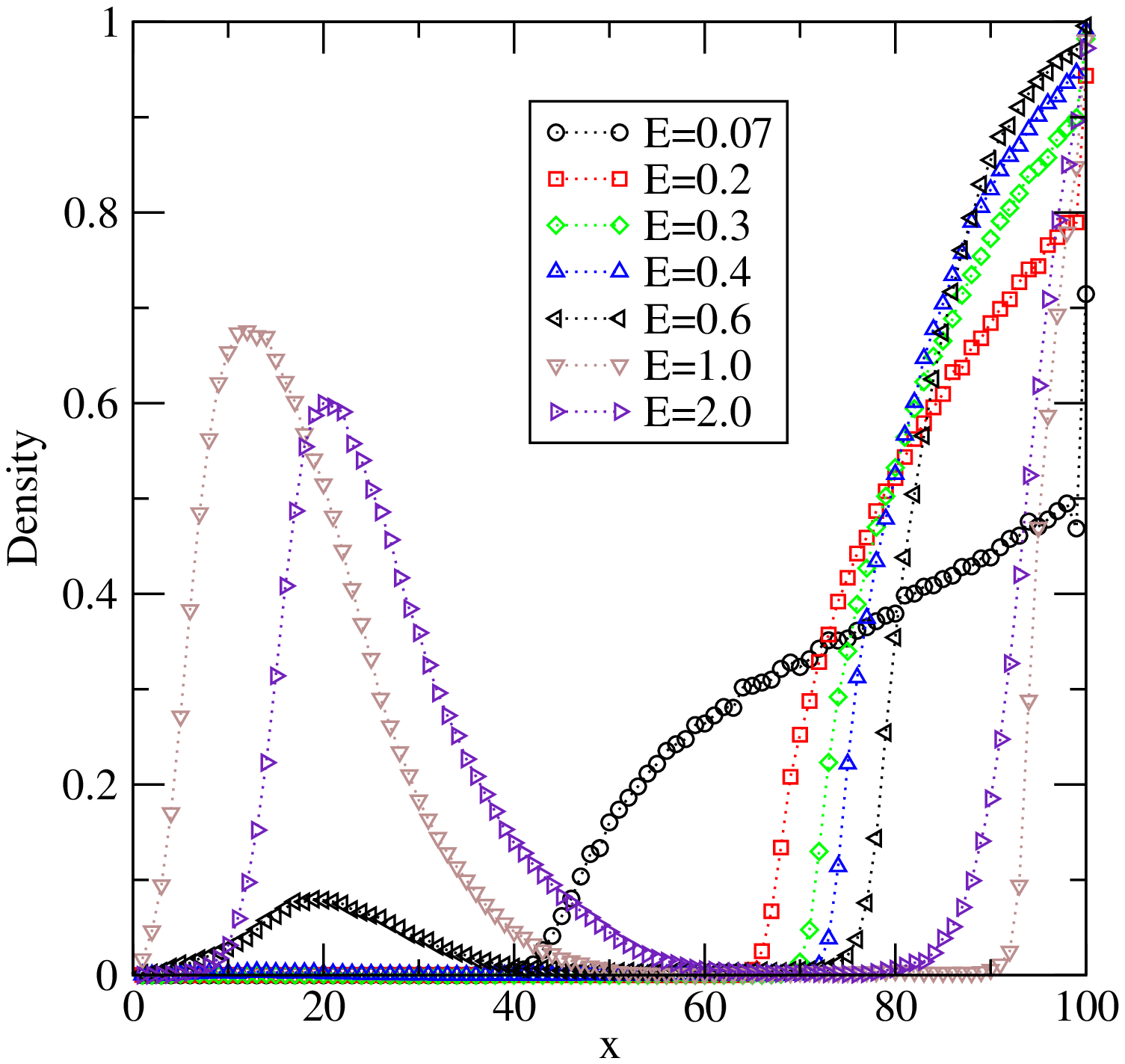}
\end{center}
\caption{
Density profiles using kink-jump ($K$) segmental 
dynamics at various fields with $L_c=39, T=1$. Ten to twenty independent 
samples were used.
}
\end{figure}

\begin{figure}[hbt]
\begin{center}
\includegraphics[angle=-00,scale=0.6]{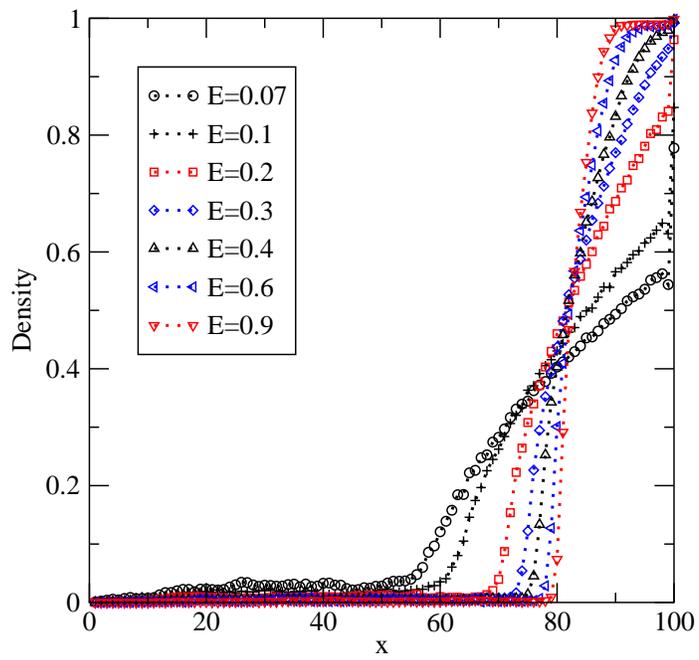}
\end{center}
\caption{
Density profiles with a combination of kink-jump and
crankshaft ($KC$) segmental dynamics.
}
\end{figure}

\begin{figure}[hbt]
\begin{center}
\includegraphics[angle=-00,scale=0.6]{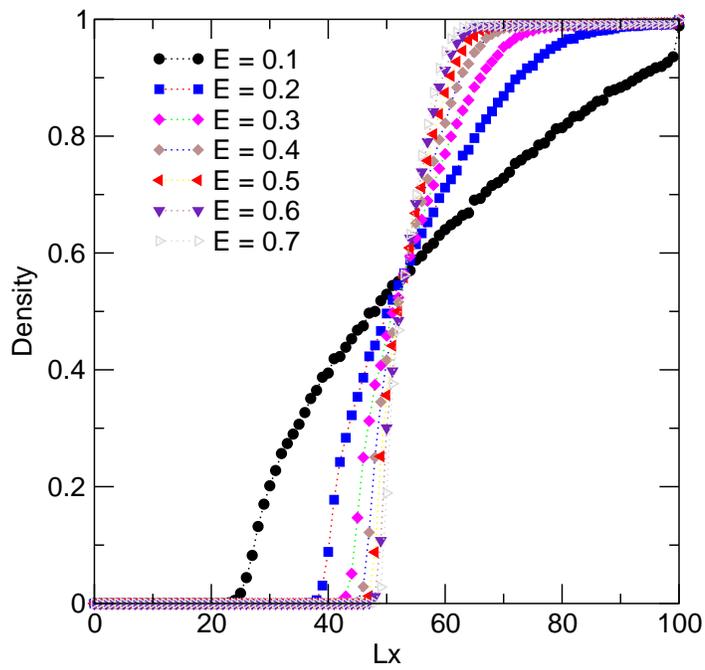}
\end{center}
\caption{
Density profiles with a combination of kink-jump,
crankshaft, and reptation ($KCR$) segmental dynamics for
$L_c=50$.
}
\end{figure}

\begin{figure}[hbt]
\begin{center}
\includegraphics[angle=-00,scale=0.6]{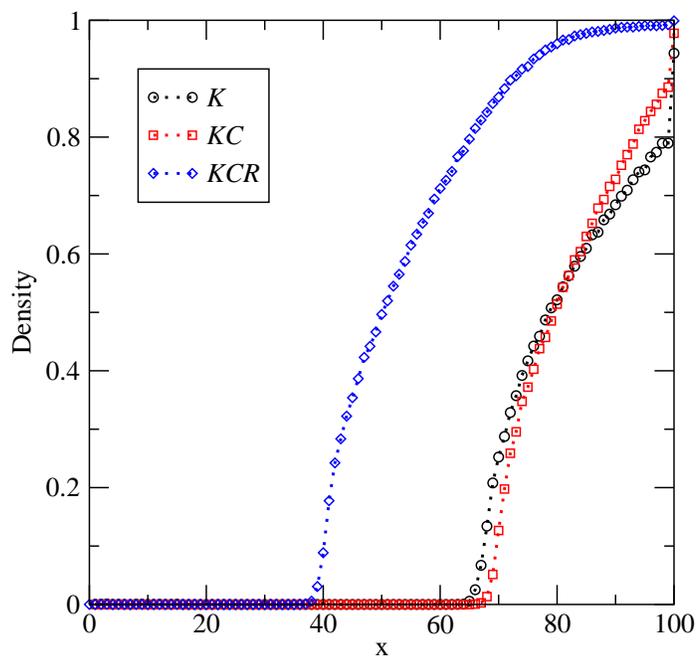}
\end{center}
\caption{
Density profiles at $T=1$, $E=0.2$ with different 
combinations of segmental dynamics, $K$, $KC$ with $L_c=39$, and
$KCR$ with $L_c=50$.
}
\end{figure}

\begin{figure}[hbt]
\begin{center}
\includegraphics[angle=-00,scale=0.6]{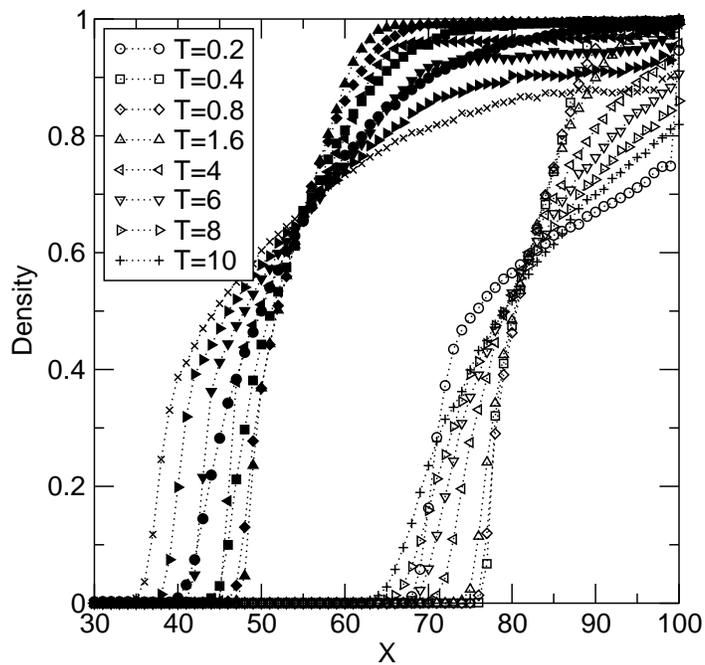}
\end{center}
\caption{
Density profiles with $KC$ $(L_c=39)$ and $KCR$ $(L_c=50)$ segmental 
dynamics at various temperatures with 
$E = 0.5$. Averages were taken over 10 independent simulation runs.
}
\end{figure}

\begin{figure}[hbt]
\begin{center}
\includegraphics[angle=-00,scale=0.6]{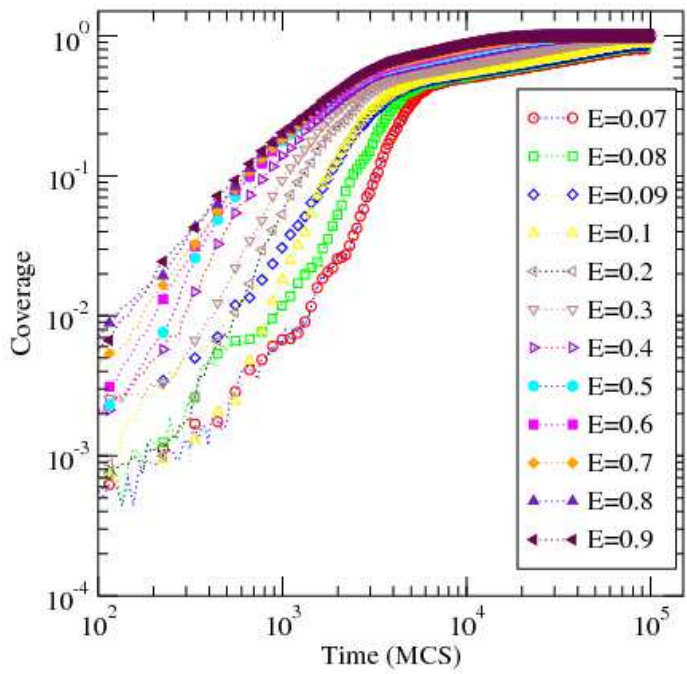}
\end{center}
\caption{
Growth of coverage with $KC$ segmental dynamics 
at various fields. Averages were taken over 10 independent simulation runs.
}
\end{figure}

\begin{figure}[hbt]
\begin{center}
\includegraphics[angle=-00,scale=0.6]{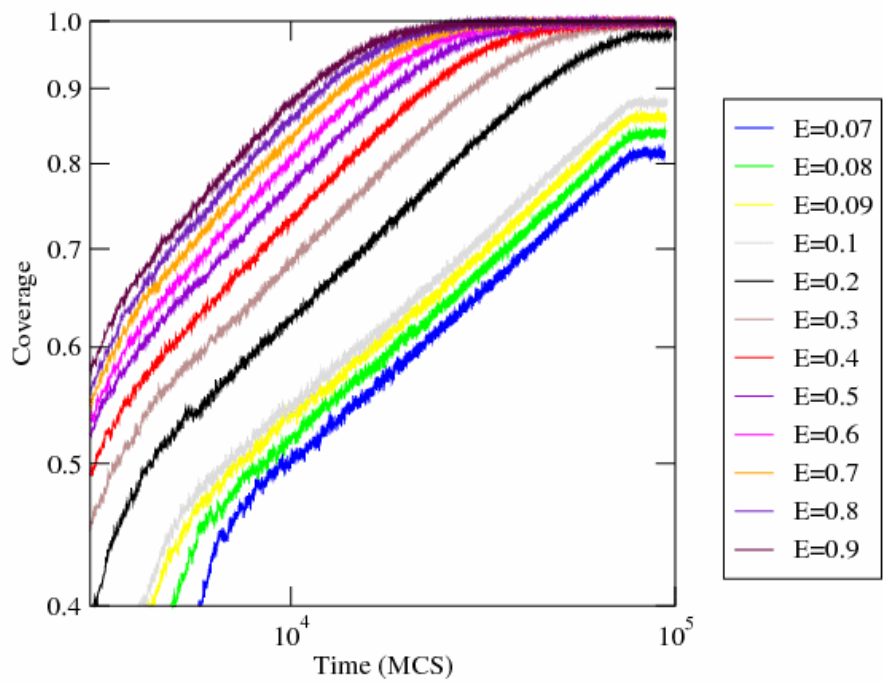}
\end{center}
\caption{
Same as Fig. 6 in a pre-saturation regions.
}
\end{figure}

\begin{figure}[hbt]
\begin{center}
\includegraphics[angle=-00,scale=0.6]{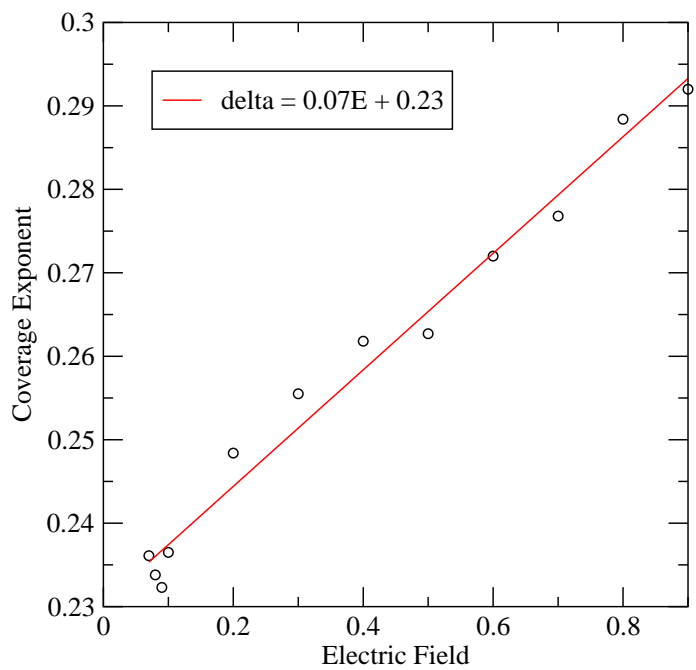}
\end{center}
\caption{
Variation of pre-saturation growth exponent
(Fig. 7) with the field.
}
\end{figure}

\begin{figure}[hbt]
\begin{center}
\includegraphics[angle=-00,scale=0.6]{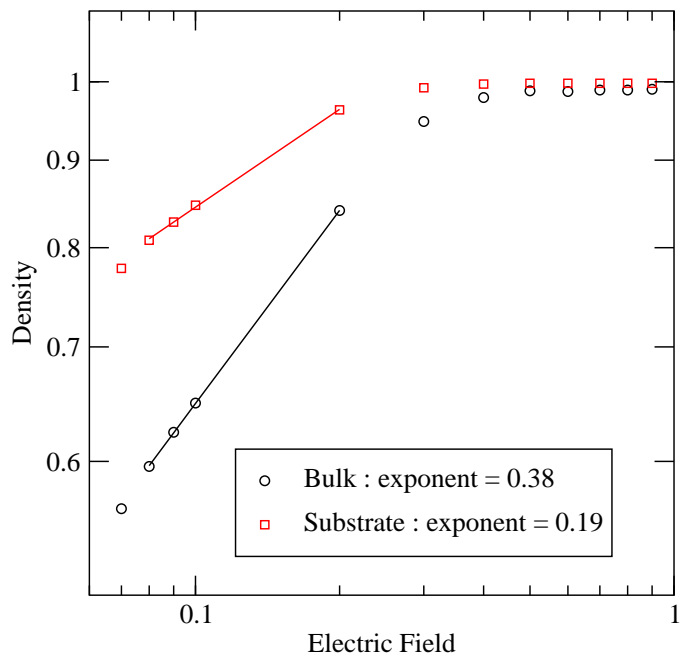}
\end{center}
\caption{
Variation of the relaxed bulk and substrate
density with the field at $T=1$ with $KC$ segmental dynamics
$L_c=39$. Averages were taken over 10 independent runs.
}
\end{figure}

\begin{figure}[hbt]
\begin{center}
\includegraphics[angle=-00,scale=0.6]{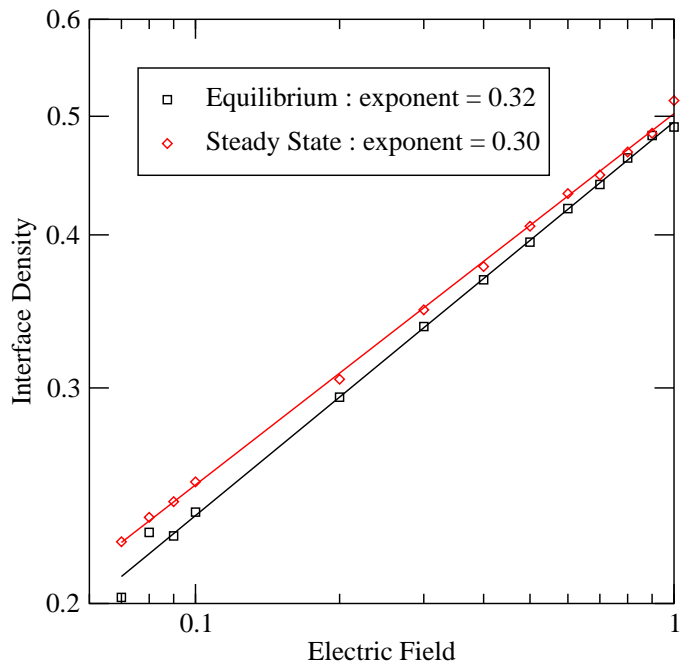}
\end{center}
\caption{
Variation of the relaxed polymer density ($d_f$)
at the interface (i.e., the front of the bulk) with the field.
Results are averaged over 10 independent runs.
}
\end{figure}

\begin{figure}[hbt]
\begin{center}
\includegraphics[angle=-00,scale=0.6]{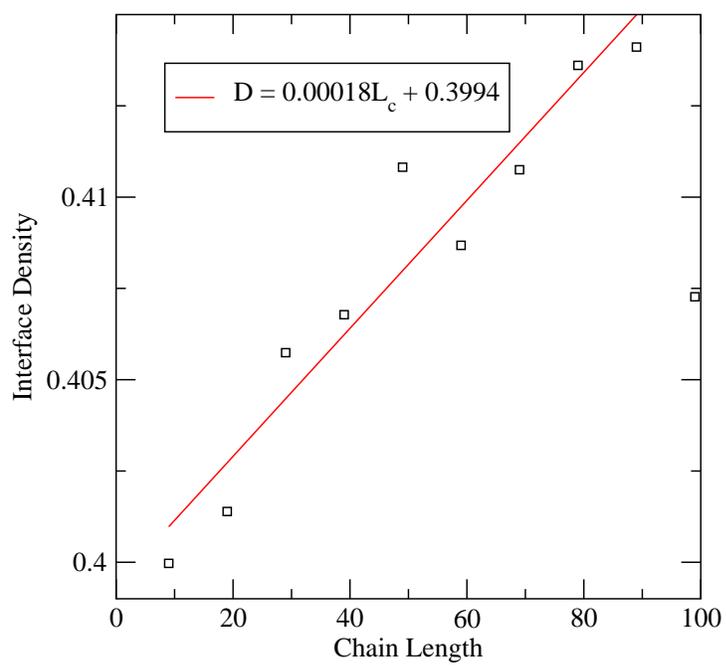}
\end{center}
\caption{
Variation of the relaxed polymer density at
the interface ($d_f$) with the chain length using $T=1, E=0.5$, and 10 
independent runs.
}
\end{figure}

\begin{figure}[hbt]
\begin{center}
\includegraphics[angle=-00,scale=0.6]{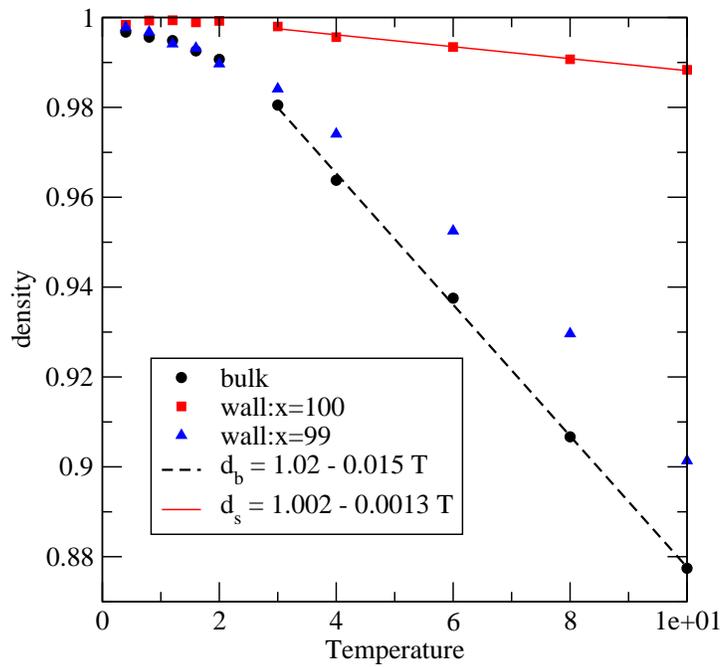}
\end{center}
\caption{
Variation of the relaxed bulk density ($d_b$), 
substrate density ($d_s$), and polymer density adjacent to substrate
($d_{as}$) with temperature at $E=0.5$ with $KCR$ segmental dynamics
($L_c = 39$).
}
\end{figure}

\begin{figure}[hbt]
\begin{center}
\includegraphics[angle=-00,scale=0.6]{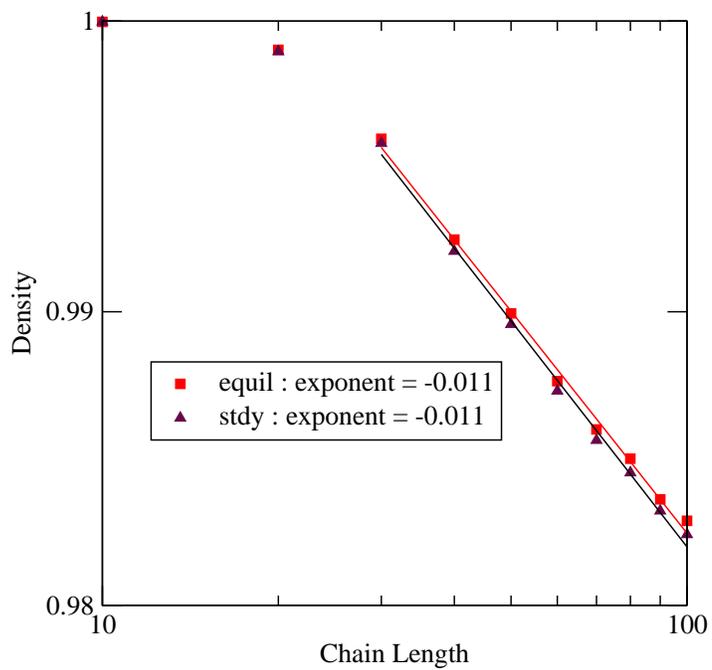}
\end{center}
\caption{
Bulk density versus chain length with $KCR$ 
segmental dynamics at $E=1.0$, $T=1.0$.
}
\end{figure}

\begin{figure}[hbt]
\begin{center}
\includegraphics[angle=-90,scale=0.6]{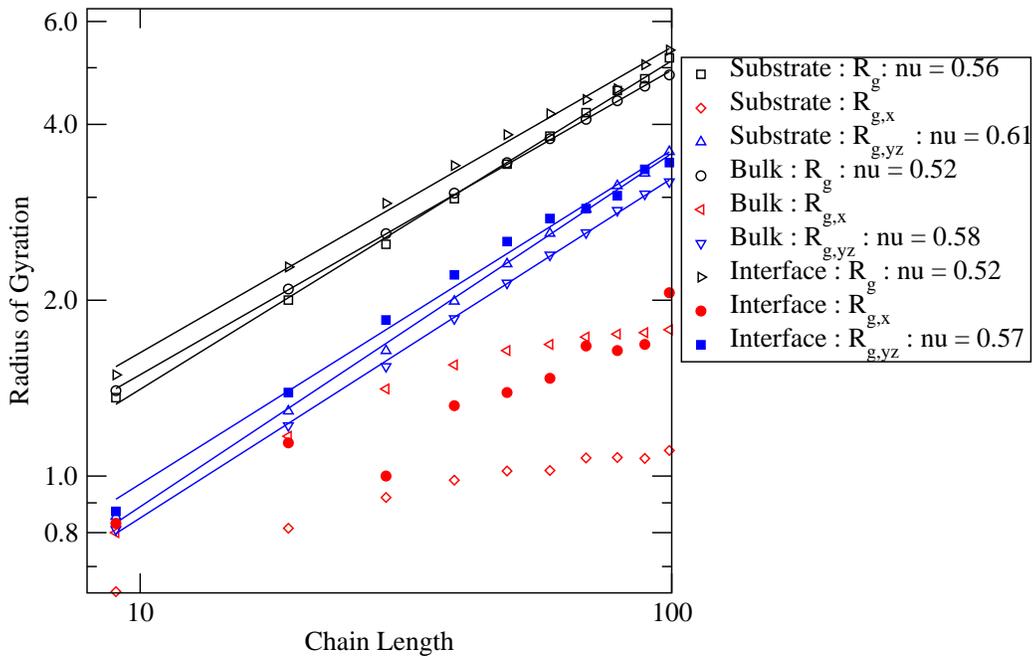}
\end{center}
\caption{
Variation of the radius of gyration of the polymer
chains $R_g$ and its longitudinal ($x$) and transverse ($yz$)
components with the molecular weight at the substrate, bulk, and
interface using 
$KC$ segmental dynamics at $T=1$, $E=0.4$.
}
\end{figure}

\end{document}